\documentclass{elsart}
\usepackage{epsfig}

\begin{document}

\begin{frontmatter}

\hbox to \textwidth{
\lower 2.5cm
\hbox to 2.5cm{
\hss
\epsfysize3cm
\epsfbox{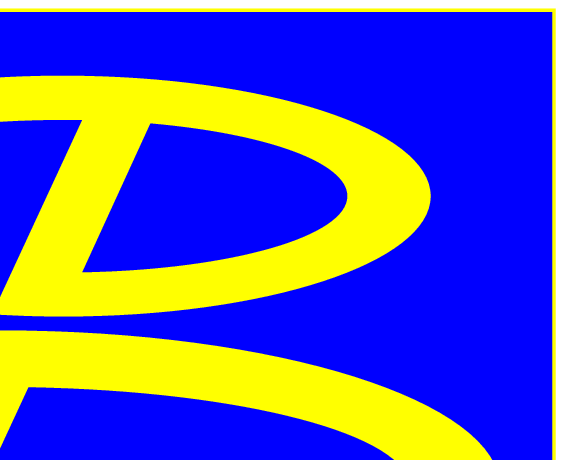}    
}
\hss
\hbox to 3cm{
\begin{tabular}{r}
{KEK preprint 2002-17} \\
{Belle preprint 2002-11}
\end{tabular}
\hss
}
}
\vspace{12pt}

\title{
Measurements of Branching Fractions and Decay Amplitudes in $B\rightarrow
J/\psi K^{*}$ decays}

\date{9 May. 2002}

\maketitle


\begin{center}
  K.~Abe$^{9}$,               
  K.~Abe$^{43}$,              
  T.~Abe$^{44}$,              
  I.~Adachi$^{9}$,            
  Byoung~Sup~Ahn$^{16}$,      
  H.~Aihara$^{45}$,           
  M.~Akatsu$^{23}$,           
  Y.~Asano$^{50}$,            
  T.~Aso$^{49}$,              
  V.~Aulchenko$^{2}$
% ...
\end{center}

\vskip 0.3cm
\begin{center}
(The Belle Collaboration)
\end{center}
\vskip 0.3cm

\small
\begin{center}
% ...
$^{2}${Budker Institute of Nuclear Physics, Novosibirsk}\\
% ...
$^{9}${High Energy Accelerator Research Organization (KEK), Tsukuba}\\
% ...
$^{16}${Korea University, Seoul}\\
% ...
$^{23}${Nagoya University, Nagoya}\\
% ...
$^{43}${Tohoku Gakuin University, Tagajo}\\
$^{44}${Tohoku University, Sendai}\\
$^{45}${University of Tokyo, Tokyo}\\
% ...
$^{49}${Toyama National College of Maritime Technology, Toyama}\\
$^{50}${University of Tsukuba, Tsukuba}\\
% ...
\end{center}

\normalsize

\begin{abstract}
The branching fractions and the decay amplitudes of
$B \rightarrow J/\psi  K^*$
decays are measured in a 29.4 fb$^{-1}$ data sample
collected with the Belle detector at the KEKB
electron-positron collider.
The decay amplitudes of helicity states of the $J/\psi K^*$ system
are determined from the full angular distribution of the final state 
particles in the transversity basis.
The branching fractions are
measured to be $(1.29\pm0.05\pm0.13) \times 10^{-3}$ for neutral
mesons
and $(1.28\pm0.07\pm0.14) \times 10^{-3}$ for charged mesons.
The measured longitudinal and transverse (perpendicular to
the transversity plane) amplitudes 
are $|A_0|^2 =
0.62\pm0.02\pm0.03$ and $|A_{\perp}|^2 = 0.19\pm0.02\pm0.03$, respectively.
The value of $|A_{\perp}|^2$ shows that the CP even component
dominates in the $B^0 \rightarrow J/\psi K^{*0}(K_S\pi^0)$ decay.

\vspace{3\parskip}

\end{abstract}


\end{frontmatter}
\clearpage

\section{Introduction}

The decay $B^0\rightarrow J/\psi  K^{*0} (K^{*0} \rightarrow K_S
\pi^0)$ has the same quark level diagram as $B^0\rightarrow
J/\psi  K_S$ and may, therefore, be used for a determination of the CP
violation parameter $\sin 2\phi_1$.
However, since the $J/\psi K^*$ system
has three possible helicity states
with different superpositions of CP eigenstates,
the dilution of the CP asymmetry depends on the degree of
polarization of the decay.
The mix of CP even and odd states must be known before a
measurement of $\sin 2\phi_1$ can be attempted.

The CP mix
can be determined through a full angular analysis of the final state
particles to obtain the decay amplitudes of the system in the
transversity basis.
Once the CP mix is measured, the CP
eigenstates of the events may be projected out in a statistical way.
The full angular analysis also serves as a test of the factorization hypothesis
by measuring the imaginary part of the decay amplitudes.

In this paper, we report the measurements of branching fractions
for $B^0 \rightarrow J/\psi  K^{*0}$ and $B^+ \rightarrow J/\psi
K^{*+}$ and the decay amplitudes of the system obtained by a full
three-dimensional  angular analysis.
The determination of sin2$\phi_1$ based on the reconstruction of
these decays is described elsewhere\cite{Hadronic}.

\section{Data sample}

The data sample used in this analysis corresponds to an integrated
luminosity of 29.4 fb$^{-1}$ recorded with the Belle detector\cite{Belle}
at the KEKB electron-positron collider\cite{KEKB}.
Events are required to satisfy the hadronic event selection 
criteria\cite{Hadronic} and have $R_2<0.5$,
where $R_2$ is the ratio of the second to zeroth Fox-Wolfram
parameters\cite{fw}.

The branching fractions and decay amplitudes
are measured by reconstructing neutral and
charged $B$ mesons in  
$B^0 \rightarrow J/\psi K^{*0}$ and $B^+ \rightarrow J/\psi K^{*+}$
(inclusion of charge conjugate modes is implied throughout this paper).
The reconstruction of the $J/\psi$ is performed using the dilepton
decays, $J/\psi \rightarrow e^+e^-$ and $\mu^+\mu^-$. For
the $e^+e^-$ mode,
electrons and positrons are identified by
matching between the energy
measured in the electromagnetic
calorimeter (ECL) and the momentum measured in the central
drift chamber (CDC), the shape of the cluster energy deposit in the
ECL, the $dE/dx$ measured in the CDC, and the light yield in the
aerogel \v{C}erenkov counters (ACC).
A likelihood is calculated
from these measurements and required to be consistent with the
electron hypothesis.
To correct for energy lost by final state radiation,  the energy of
any cluster in the ECL within 0.05 radians of the track momentum is added
to that of the track.
The invariant mass of each pair of identified electrons is
calculated, and the pair is identified as  $J/\psi \rightarrow
e^+e^-$ if the
mass is in the range $2.95~{\rm GeV}/c^2 < M(e^+e^-) < 3.15~{\rm
GeV}/c^2$.
Tracks are identified as muons 
by means of a likelihood number based on
$(i)$ a comparison of the number of layers with associated hits in
the muon detector (KLM) with the  number expected based on momentum and
$(ii)$ the energy of the associated hit in the ECL.
An oppositely charged pair of identified muons  is identified as
$J/\psi \rightarrow \mu^+\mu^-$ if the invariant mass is in the range
$3.05~{\rm GeV}/c^2 < M(\mu^+\mu^-) < 3.15~{\rm GeV}/c^2$.
To improve the momentum resolution, a kinematic fit which uses the
$J/\psi$ mass as a constraint
is performed on $J/\psi$ candidates passing the above selections.

$K^{*0}$ and  $K^{*+}$ candidates
are reconstructed in the decay modes
$K^{*0}\rightarrow K^+\pi^-$, $K^{*+}\rightarrow K_S\pi^+$,
$K^{*+}\rightarrow K^+\pi^0$, and $K^{*0}\rightarrow
K_S\pi^0$.
Charged kaons are identified by requiring the kaon likelihood of a
track to be consistent with expectations.
The kaon likelihood is
obtained by combining measurements of the time of flight measured by the 
scintillation counters (TOF),
$dE/dx$ by the CDC and the hit information in the ACC.
Tracks which are not identified
as kaons and not used as leptons in the $J/\psi$ reconstruction
are treated as charged pion candidates.
$K_S$ candidates are
reconstructed from pairs of oppositely charged tracks which satisfy
three conditions: 1) the
distance of the closest approach of both 
tracks to the nominal interaction point is larger than 0.03 cm,
2) the angle between the $K_S$ momentum vector and the vector
displacement of the $K_S$ vertex point from the $J/\psi$ vertex is
less than 0.15 radians, and
3) the reconstructed decay vertex of the $K_S$ is at least 0.1 cm
away from the interaction point.
Each pair with  invariant mass satisfying
$0.47~{\rm GeV}/c^2 < M(\pi^+\pi^-) < 0.52~
{\rm GeV}/c^2$ is identified as a $K_S$.
To improve the mass resolution the momenta are re-fitted,
constraining both tracks to originate at the reconstructed vertex.
Candidate $\pi^0$ mesons are reconstructed from
clusters in the ECL
that are unmatched to charged tracks and have energy greater than
$40~{\rm MeV}$.
A photon pair with an invariant mass
in the range
$0.125~{\rm GeV}/c^2 < M(\gamma\gamma) < 0.145~{\rm GeV}/c^2$ is identified
as a $\pi^0$. A mass-constrained fit is performed to obtain the
momentum of the $\pi^0$.
A $K^{*}$ is identified if the absolute difference between
the invariant mass of
an identified $K \pi$ pair and the nominal $K^*(892)$ mass
is less than $75~{\rm MeV}/c^2$.

\section{Measurement of branching fractions}

Candidate $B^0$ and $B^{+}$ mesons are reconstructed by selecting
events with a $J/\psi$ and a $K^*$ 
and examining two
quantities in the center-of-mass of the $\Upsilon$(4S), the
beam-constrained mass ($M_{bc}$)
and the energy difference between the $B$
candidate and the beam energy ($\Delta E$).
The beam-constrained mass, which is the invariant mass of a reconstructed
$J/\psi$ and $K^*$ calculated taking the energy to be the beam
energy, is required to be
in the range $5.20-5.29~{\rm GeV}/c^2$.
For modes with charged tracks only,
$|\Delta E|$ is required to be less than $30~{\rm MeV}$.
For decay modes that include a $\pi^0$,
$\Delta E$ is required to satisfy $-50~{\rm MeV} <
\Delta E < 30~{\rm MeV}$.
To eliminate slow $\pi^0$ backgrounds,
the angle of the kaon with respect
to the $K^*$ direction in the  $K^*$ rest frame, $\theta_{K^*}$, is required
to satisfy ${\rm cos}\theta_{K^*}<0.8$. This is equivalent to a
requirement that 
the $\pi^0$ momentum be greater than $175~{\rm MeV}/c$.
When an event contains more than one candidate passing the above requirements,
the combination for which $\Delta E$ is closest to zero is selected.
Defining the signal region as $5.27~{\rm GeV}/c^2<M_{bc}<5.29~{\rm GeV}/c^2$,
we find that the numbers of events passing the selection
criteria and lying in the signal region are 968 for $J/\psi
K^{*0}(K^+\pi^-)$, 241 for $J/\psi K^{*+}(K_S\pi^+)$, 220 for $J/\psi
K^{*+}(K^+\pi^0)$ and 42 for
$J/\psi K^{*0}(K_S\pi^0)$.

\begin{figure}
\centerline{\mbox{\psfig{figure=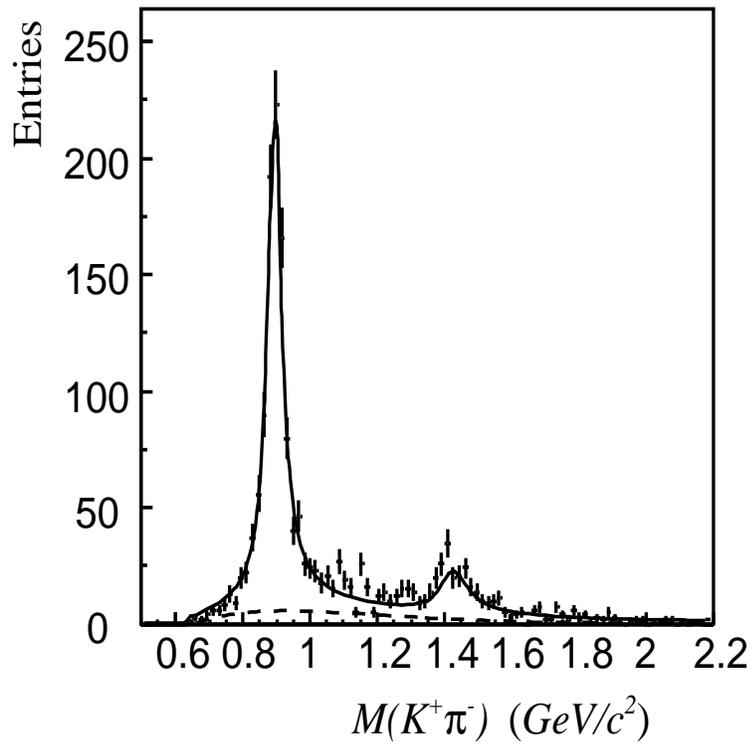,width=10cm,height=10cm}}}
\caption{ The invariant $K \pi$ mass distribution for 
$B^0 \rightarrow J/\psi K^{*0}(K^+\pi^-)$
candidates. The solid line shows a
fit to two Breit-Wigner functions corresponding to $K^*(892)$ and
$K^*_2(1430)$ with a background function (dashed line). \label{kstar}}
\end{figure}

\begin{figure}
\centerline{\epsfysize=10cm \epsfbox{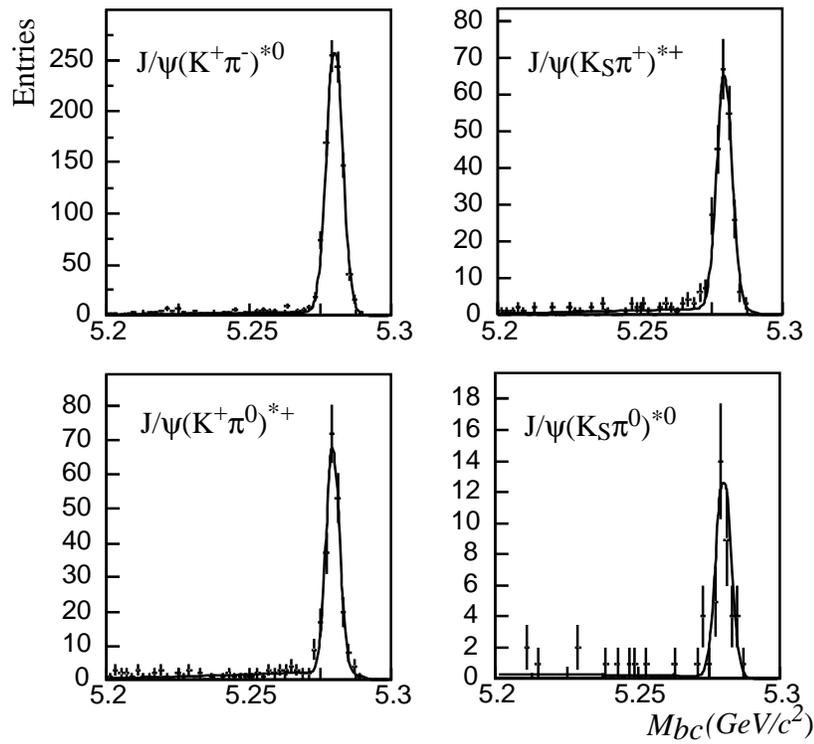}}
\caption{The distributions of the beam-constrained mass for four
reconstructed modes. The solid lines show the fits.\label{mbcfit}}
\end{figure}

For each reconstructed mode, the distribution in $M_{bc}$ is binned and
fitted to a parameterized function of the form
\begin{equation}
N(M_{bc}) = f_{sig}(M_{bc}) +
\sum_{i}f_{cf}^{i}(M_{bc}) + f_{nr}(M_{bc}) + f_{combi}(M_{bc}),
\end{equation}
where $f_{sig}$, $f_{cf}^i$, $f_{nr}$, and $f_{combi}$ are the shapes
for the signal, background from other $B\rightarrow J/\psi K^*$ modes
(which we will refer to as cross-feed), background from non-resonant
$B
\rightarrow J/\psi K \pi$, and combinatorial background, respectively.
$f_{sig}$ is a Gaussian with the peak position fixed at the $B$ mass and
a width that is determined from Monte Carlo.
Only the amplitude is  varied in fitting.
$f_{cf}^i$ is described by a separate Gaussian for each of the three
$J/\psi K^*$ decay modes other than the signal mode.
Monte Carlo simulations are used to determine the mean and width for
each mode, as well as the efficiency for it to pass the signal mode
criteria.
The amplitudes of $f_{cf}^i$ are scaled such that the ratios of
cross-feed to signal efficiency are fixed
assuming equal production rates and 
decay branching fractions for the neutral and charged $B$'s.
The rate of cross-feed from $B \rightarrow J/\psi K$ modes is also
estimated using Monte Carlo and found to be negligible.
The total amount of the cross-feed
is estimated to be 1.8--6.2\% (depending on the mode) 
of the events in the signal region.
The shape of $f_{nr}$ is also assumed to be a Gaussian.
While the mean and width are expected to be similar to those of
$f_{sig}$, they are determined separately to allow for minor
kinematic differences,
using events in the $K^*$ mass sideband ($1.1~{\rm
GeV}/c^2<M(K\pi)<1.3~{\rm GeV}/c^2$).
The magnitude of the non-resonant contribution is derived from fitting the
$K\pi$ mass distribution for events
selected by the criteria used for the signal reconstruction without
requiring the $K\pi$ mass to be inside
the $K^*$ mass window.
The $K\pi$ mass distribution (Fig.~\ref{kstar}) is fitted
with P-wave and D-wave Breit-Wigner functions describing the $K^*(892)$ and
$K^*_2(1430)$ mass peaks, respectively, and
a threshold function constrained at the mass threshold
to describe the non-resonant production.
The peak positions and widths of the Breit-Wigner functions are fixed to
the nominal PDG values\cite{PDG}.
The amplitude of $f_{nr}$ is determined from the contribution of the
fitted background functions in the $K^*$ mass that takes into account 
the contamination of the combinatorial background in the functions.
The estimated contribution of the non-resonant decay in the signal
region is 5.2\%.
$f_{combi}$ is the line shape for the combinatorial background and is
modeled by
the ``ARGUS function''\cite{ARGUS}, where the parameters are free in
the fit. 
Fig.~\ref{mbcfit} shows the results of the fits to four modes.
The fits indicate the contamination of the combinatorial background is
less than 1\% in the signal region.

The branching fraction is calculated from the yield in the
signal region obtained from the fitted $f_{sig}$.
The detection efficiency for each mode is estimated by applying
the selection criteria to Monte Carlo event samples.
Events are generated unpolarized
in the $J/\psi K^*$ system
using the $QQ$ event generator\cite{QQ} and
are passed through a detector simulation program based on
GEANT3\cite{GEANT}.
The estimated efficiencies
are 28.4\%, 20.7\%, 11.8\%, and 8.1\%,  for  $J/\psi
K^{*0}(K^+\pi^-)$, $J/\psi K^{*+}(K_S\pi^+)$, $J/\psi
K^{*+}(K^+\pi^0)$ and $J/\psi K^{*0}(K_S\pi^0)$, respectively.
The branching fractions of secondary decays
are fixed to the PDG
values\cite{PDG} and the numbers of neutral and charged
$B\overline{B}$ pairs from $\Upsilon(4S)$ decays are assumed to be equal.

The systematic uncertainties on the branching fractions are dominated by the
following sources: 1) tracking efficiency (2.0\% per
track), 2) $\pi^0$ reconstruction (3.0\%), 3) lepton identification
efficiency (5.0\%), 4) kaon identification efficiency ($<$ 1.0\%),
5) background estimation (2.3--4.8\% depending on
the mode), 6) number of
$B\overline{B}$ events (1.0\%), 7) polarization (1.7\%), and 8)
branching fraction for secondary decays (1.7\%).
The estimate of  the non-resonant $B \rightarrow J/\psi K \pi$
background has a relatively large uncertainty because the fit to the
$K\pi$  mass distribution is poor in the region of $K\pi$ mass from
1.1~GeV/$c$ to 1.3~GeV/$c$, where some excess is observed.
This excess has also been seen elsewhere\cite{BaBar,CLEO-b}, and
there has been some speculation as to its origin\cite{BaBar}.
As its source is not established, we estimate the uncertainty in its
contribution to the signal by including in the fit to the $K\pi$ mass
distribution an additional term that is constrained to zero at the
mass threshold and extends linearly to 1.35~GeV/$c^2$.
The change to the signal yield due to this modification is taken to
be the uncertainty.

The resulting branching fractions are summarized in
Table~\ref{brlist} with the estimated systematic errors.

\section{Measurement of decay amplitudes}

The decay amplitudes of $B\rightarrow J/\psi K^*$ decays are
measured in the transversity basis\cite{transversity}.
The direction of motion of the $J/\psi$ in the rest frame of the $B$
candidate is defined to be the $x$-axis. The $y$-axis is defined to be
in the direction of the projection of the $K$ momentum into the plane
perpendicular to the $x$-axis in the $B$ rest frame. The $x$-$y$ plane
contains the momenta of the $J/\psi$, the $K$, and the $\pi$.
The $z$-axis is defined to be perpendicular to the $x$-$y$ plane
according to the right-hand rule.
The angle between the $l^+$ direction and the
$z$-axis 
in the $J/\psi$ rest frame is defined as $\theta_{tr}$.
The angle between the $x$-axis and the projection
of the $l^+$ momentum
onto the $x$-$y$ plane is defined as $\phi_{tr}$ in the same frame. 
The angle $\theta_{K^*}$ is defined as described in the previous section.

The distribution of these three
angles for $B \rightarrow J/\psi  K^*$ decays
is described in terms of three amplitudes\cite{Yamamoto}:
\begin{eqnarray}
\frac{1}{\Gamma}\frac{d\Gamma}{d\cos\theta_{tr}d\cos\theta_{K^*}d\phi_{tr}}
& = &    \frac{9}{32\pi}\times
     [ 2\cos^2\theta_{K^*}(1-\sin^2\theta_{tr}\cos^2\phi_{tr})|A_0|^2
\nonumber \\
& &   + \sin^2\theta_{K^*}(1-\sin^2\theta_{tr}\sin^2\phi_{tr})|A_{\parallel}|^2
\nonumber \\
& &   + \sin^2\theta_{K^*}\sin^2\theta_{tr}|A_{\perp}|^2 \nonumber \\
& &   + \eta \sin^2\theta_{K^*}\sin 2\theta_{tr} \sin\phi_{tr} 
Im(A_{\parallel}^*A_{\perp}) \nonumber \\
& &   - \frac{1}{\sqrt{2}} \sin 2\theta_{K^*} \sin^2\theta_{tr} \sin 
2\phi_{tr} Re(A_0^*A_{\parallel}) \nonumber \\
& &   + \eta \frac{1}{\sqrt{2}} \sin 2\theta_{K^*} \sin 2\theta_{tr}
\cos\phi_{tr} Im(A_0^*A_{\perp}) ], \label{theory}
\end{eqnarray}
where $A_0$, $A_{\parallel}$ and $A_{\perp}$ are the complex
amplitudes of the three helicity states in the trasversity basis, and
$\eta = +1~(-1)$ for $B^0$ and $B^+$ ($\overline{B^0}$  and $B^-$).
$|A_0|^2$ denotes the longitudinal polarization of $J/\psi$ while
$|A_{\perp}|^2 (|A_{\parallel}|^2)$ is the transverse polarization
component along the z-axis (y-axis).
Also, $|A_0|^2 + |A_{\parallel}|^2$ is the amplitude corresponding to
the CP-even state, while $|A_{\perp}|^2$ is the CP-odd component in
$B^0 \rightarrow J/\psi K^{*0}(K^{*0}\rightarrow K_S  \pi^0)$.
These amplitudes are normalized so that:
\begin{equation}
|A_0|^2 + |A_{\parallel}|^2 + |A_{\perp}|^2 = 1. \label{NORM}
\end{equation}

The amplitudes are determined by fitting this function to the
measured three-dimensional distribution in $\theta_{tr}$, $\phi_{tr}$
and $\theta_{K^*}$, taking into account the detection efficiency and 
background.
The resolution of the angular measurements is estimated by Monte Carlo
and found to be typically less than 0.02 radians.
The value of $\eta$ is determined from the charge of pions
or kaons used in the $K^*$ reconstruction.
We do not
include the $B^0\rightarrow
J/\psi K^{*0}(K_S \pi^0)$
mode in the fit since in this case $\eta$ is
not well-defined.

The fit is performed using an unbinned maximum likelihood
method. The probability density function (PDF) is defined using the
theoretical distribution in (\ref{theory}) and can be expressed as
\begin{eqnarray}
PDF (x,y,z, M_{bc}) & = & N \times
  [ f_{sig}(M_{bc}) \times \epsilon(x,y,z) \times \frac{1}{\Gamma}
\frac{d^3\Gamma}{dxdydz}(x,y,z) \nonumber \\
& & +~ \sum_i f_{cf}^i(M_{bc}) \times ADF_{cf}(x,y,z)\nonumber \\
& & +~ f_{nr}(M_{bc}) \times ADF_{nr}(x,y,z) \nonumber \\
& & +~ f_{combi}(M_{bc}) \times ADF_{combi}(x,y,z) ], \label{PDF1}
\end{eqnarray}
where $x = {\rm cos}\theta_{tr}$, $y = \phi_{tr}$, and $z = {\rm
cos}\theta_{K^*}$, $N$ is the normalization factor of the PDF,
$\epsilon(x,y,z)$ is the detection efficiency as a function of the three 
angles, and
$ADF_{cf}$, $ADF_{nr}$, $ADF_{combi}$ are the angular shapes for the
cross-feed, non-resonant and combinatorial backgrounds, respectively.

Here, $f_{sig}(M_{bc})$, $f_{cf}^i(M_{bc})$, $f_{nr}(M_{bc})$, and
$f_{combi}(M_{bc})$ are the fractions of
signal events, cross-feed contamination, non-resonant
contamination and combinatorial background as a function of beam
constrained mass,
respectively. These fractions
are obtained from the measurements of the branching
fractions described in the previous section.

The detection efficiency function $\epsilon(x,y,z)$ is
obtained from a large Monte Carlo sample of 1 million events for each
mode generated without any polarization of the
$B\rightarrow J/\psi K^*$ system.
Events are histogrammed in a $20\times20\times20$ grid in a
${\rm cos}\theta_{tr}-\phi_{tr}-{\rm cos}\theta_{K^*}$
cube. The distribution is fitted to a three-dimensional polynomial 
with correlations taken into account.
The efficiency is almost flat
except in the region $\cos\theta_{K^*} \sim 1$,
where the pion is slow so that the efficiency is reduced.

The angular distribution function for the cross-feed background
($ADF_{cf}$) is
determined from the distribution in Monte Carlo events.
The function for the non-resonant production ($ADF_{nr}$) is
determined from
events in the sideband of the $K\pi$ mass distribution.
The distribution for the combinatorial background ($ADF_{combi}$) is
obtained from events in the sideband of the beam-constrained
mass.
These distributions are parameterized as polynomials where the
parameters are determined from the fit.

\begin{figure}
\centerline{\mbox{\psfig{figure=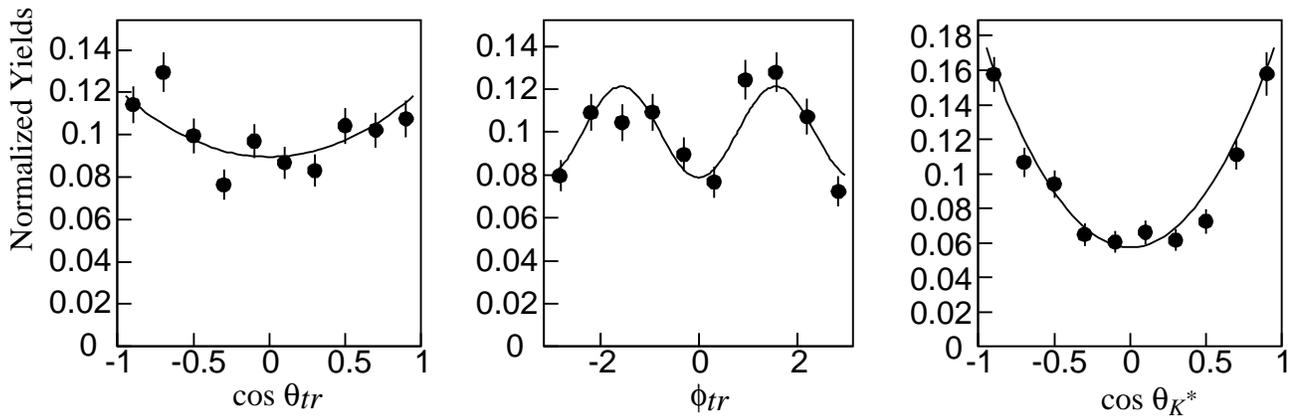,width=5.5cm,angle=270}}}
\caption{Projected distributions of $\cos\theta_{tr}$,
$\phi_{tr}$ and $\cos\theta_{K^*}$ for the combined data. The distributions
are background subtracted and acceptance corrected. The solid line
shows the theoretical predictions with the obtained decay amplitudes.
\label{angfinal}}
\end{figure}

In the fit to the angular distributions,
the imaginary part of $A_0$ is defined to be zero
relative to the imaginary parts of the other amplitudes since the
overall phase of the decay amplitudes is arbitrary.
By applying the normalization condition (\ref{NORM}), four
parameters, $|A_0|^2$, $|A_{\perp}|^2$, $arg(A_{\parallel})$ and
$arg(A_{\perp})$ are left be determined from the fit. The normalization of
the PDF ($N$) is calculated by numerical integration over the three
dimensional angular cube whenever parameter values are changed.
A likelihood is defined as the product of the PDF for all the
events and the values of parameters are determined by maximizing the
likelihood.
We perform separate fits to the three decay modes as well as
a combined fit by defining a single likelihood.
The parameter values determined from these fits are summarized in
Table~\ref{fitres}.
The projected angular distributions of the data for all $K^*$ modes
combined are shown in
Fig.~\ref{angfinal}. The distributions are corrected for the effects
of detector acceptance and backgrounds.

Systematics in the fit are studied for the following uncertainties:
1) efficiency function (Monte Carlo statistics and effect of polarization), 2)
angular distribution functions for backgrounds, 3) background
fractions, 4) slow pion efficiency, and 5) fit algorithm.
These contributions to the systematic error are
summarized in Table~\ref{angsyse}. The dominant contributions arise
from the dependence of the efficiency and background functions on the
polarization, 
and also the uncertainty in the detection efficiency
for slow pions. The
first uncertainty is estimated by comparing the functions obtained for
Monte Carlo samples generated with and without the polarization
in the $J/\psi K^*$ system. The second uncertainty dominates in the
$\cos \theta_{K^*}$ distribution where $\cos \theta_{K^*}$ is close to
1. The uncertainty is estimated by comparing with
the results with a cut at $\cos \theta_{K^*} <0.9$ for decay modes with
$\pi^{\pm}$ and with a cut at $\cos \theta_{K^*}<0.7$ for the mode
with $\pi^0$. The systematics in the fit algorithm are studied using
toy Monte Carlo events. The differences between input decay amplitudes
for Monte Carlo and the values obtained by the fit are
taken as the uncertainties.

\section{Conclusion}
Branching fractions for
$B^{0}\rightarrow J/\psi K^{*0}$ and $B^{+}\rightarrow J/\psi K^{*+}$
are measured via reconstruction of $K^{*0}\rightarrow K^+ \pi^-$, 
$K^{*0}\rightarrow K_S\pi^0$, $K^{*+}\rightarrow K^+ \pi^0$, and 
$K^{*+}\rightarrow K_S \pi^+$  with leptonic decays of $J/\psi$.
The resultant branching fractions are  ${\it Br}(B^0 \rightarrow
J/\psi  K^{*0}) =
( 1.29 \pm 0.05 \pm 0.13 ) \times 10^{-3}$ and
${\it Br}(B^+ \rightarrow J/\psi K^{*+})  =
( 1.28 \pm 0.07 \pm 0.14 ) \times 10^{-3}$.
These values are consistent with the measurements by
CLEO\cite{CLEO-b}and BaBar\cite{BaBar-Br} (Table ~\ref{Br-others}).

The decay amplitudes for $B\rightarrow
J/\psi K^*$ are measured
by fitting the angular distribution of
final state particles in the transversity basis to a theoretical
distribution with the effects of the detector acceptance and the
background taken into account. In the fit, the sum of amplitudes is
normalized to 1 ($|A_0|^2+|A_\parallel|^2+|A_\perp|^2=1$) and
$arg(A_0)$ is defined to be zero. The fit is performed using an unbinned
maximum likelihood method. From the fit, the values obtained are
$|A_0|^2  =  0.618 \pm 0.020 \pm 0.027$,
$|A_\perp|^2  =  0.191 \pm 0.023 \pm 0.026$,
$arg(A_{\parallel})  =  2.83 \pm 0.19 \pm 0.08$, and
$arg(A_{\perp}) = -0.09 \pm 0.13 \pm 0.06$.
The measured value of $|A_{\perp}|^2$ shows that the CP even component
dominates in the decay $B^0\rightarrow J/\psi K^{*0}(K_S\pi^0)$. The
value is used in the determination of $\sin 2\phi_1$ as
described in \cite{Hadronic}.
Table ~\ref{others} shows the comparison with other measurements.
The parameter $arg(A_{\parallel})$ is sensitive to final state 
interactions (FSI).
A shift from 0 or $\pi$ shows the existence of the FSI. 
The measured value is consistent with results from
other experiments\cite{BaBar}\cite{CDF}. However, the significance of
its deviation from $\pi$ is not sufficient to conclude the existence
of FSI.

We wish to thank the KEKB accelerator group for the excellent
operation of the KEKB accelerator.
We acknowledge support from the Ministry of Education,
Culture, Sports, Science, and Technology of Japan
and the Japan Society for the Promotion of Science;
the Australian Research Council
and the Australian Department of Industry, Science and Resources;
the National Science Foundation of China under contract No.~10175071;
the Department of Science and Technology of India;
the BK21 program of the Ministry of Education of Korea
and the CHEP SRC program of the Korea Science and Engineering
Foundation;
the Polish State Committee for Scientific Research
under contract No.~2P03B 17017;
the Ministry of Science and Technology of the Russian Federation;
the Ministry of Education, Science and Sport of the Republic of
Slovenia;
the National Science Council and the Ministry of Education of Taiwan;
and the U.S.\ Department of Energy.

\clearpage

%
%

\clearpage

%
%

%
%
\begin{table}
\begin{center}
\caption{Measured branching fractions for $B^0\rightarrow
J/\psi K^{*0}$ and $B^+\rightarrow J/\psi K^{*+}$. \label{brlist}}
\begin{tabular}{|c|c|}
\hline
Mode & Branching Fraction ($\times 10^{-3}$) \\ \hline
$B^0\rightarrow J/\psi   K^{*0}(K^+\pi^-)$ & $1.30 \pm 0.05 \pm 0.12 $ \\
$B^0\rightarrow J/\psi   K^{*0}(K_S\pi^0)$ & $1.20 \pm 0.20 \pm 0.15 
$ \\ \hline
$B^0\rightarrow J/\psi   K^{*0}$ & $1.29 \pm 0.05 \pm 0.13$ \\ \hline
$B^+\rightarrow J/\psi   K^{*+}(K_S\pi^+)$ & $1.23 \pm 0.11 \pm 0.15 $ \\
$B^+\rightarrow J/\psi   K^{*+}(K^+\pi^0)$ & $1.33 \pm 0.10 \pm 0.14 
$ \\ \hline
$B^+\rightarrow J/\psi   K^{*+}$ & $1.28 \pm 0.07 \pm 0.14$ \\
\hline
\end{tabular}
\end{center}
\end{table}

%
%
\begin{table}
\begin{center}
\caption{Measured decay amplitudes for $B^0\rightarrow
J/\psi K^{*0}$ and $B^+\rightarrow J/\psi K^{*+}$ (statistical error only).
\label{fitres}}
\begin{tabular}{|c|c|c|c|c|}
\hline
Mode & $|A_0|^2$ & $|A_{\perp}|^2$ & $arg(A_{\parallel})$ & $arg(A_{\perp})$
\\ \hline
$B^0\rightarrow J/\psi   K^{*0}(K^+\pi^-)$ &
$0.607\pm0.020$ & $0.195\pm0.023$ & $2.87\pm0.19$ & $0.04\pm 0.13$ \\
$B^+\rightarrow J/\psi   K^{*+}(K_S\pi^+)$ &
$0.630\pm0.048$ & $0.171\pm0.057$ & $2.74\pm0.39$ & $-0.45\pm 0.33$ \\
$B^+\rightarrow J/\psi   K^{*+}(K^+\pi^0)$ &
$0.653\pm0.052$ & $0.194\pm0.087$ & $2.69\pm1.08$ & $-0.23\pm 0.56$ \\ \hline
Combined &
$0.617\pm0.020$ & $0.192\pm0.023$ & $2.83\pm0.19$ & $-0.09\pm 0.13$ \\
\hline
\end{tabular}
\end{center}
\end{table}

%
%
\begin{table}
\begin{center}
\caption{Systematic errors in the measurement of decay amplitudes.
  \label{angsyse}}
\begin{tabular}{|c|c|c|c|c|}
\hline
Item & $|A_0|^2$ & $|A_\perp|^2$ & $arg(A_{\parallel})$ &
$arg({A_\perp})$ \\ \hline
Efficiency          & 0.008 & 0.003 & 0.04 & 0.03 \\
ADF for backgrounds & 0.007 & 0.003 & 0.01 & 0.04 \\
Background fractions & 0.001 & 0.001 & 0.01 & 0.01 \\
Slow pion efficiency & 0.025 & 0.026 & 0.07 & 0.02 \\
Fit algorithm        & 0.001 & 0.001 & $<$0.01 & 0.01 \\
\hline
Total & 0.027 & 0.026 & 0.08 & 0.06 \\
\hline
\end{tabular}
\end{center}
\end{table}

%
%
\begin{table}
\begin{center}
\caption{Comparison with previous measurements of branching fractions for
$B^0\rightarrow J/\psi K^{*0}$ and $B^+\rightarrow J/\psi K^{*+}$. The 
first (second) quoted error is statistical (systematic).
\label{Br-others}}
\begin{tabular}{|c|c|c|}
\hline
  & ${\it Br}(B^0\rightarrow J/\psi K^{*0})\;(\times 10^{-3})$ & 
${\it Br}(B^+\rightarrow
  J/\psi K^{*+})\;(\times 10^{-3})$ \\ \hline
CLEO\cite{CLEO-b} & $1.32 \pm 0.15 \pm 0.17$ & $1.41 \pm 0.20 \pm 0.24$ \\
BaBar\cite{BaBar-Br} & $1.24 \pm 0.05 \pm 0.09$ & $1.37 \pm 0.09 \pm
  0.11$ \\ \hline
This measurement & $1.29 \pm 0.05 \pm 0.13$ & $1.28 \pm 0.07 \pm 0.14$
  \\
\hline
\end{tabular}
\end{center}
\end{table}

%
%
\begin{table}
\begin{center}
\caption{Comparison with previous measurements of decay amplitudes in
$B\rightarrow J/\psi   K^*$ . The first (second) quoted error
is statistical (systematic). \label{others}}
\begin{tabular}{|c|c|c|c|c|}
\hline
Group & $|A_0|^2$ & $|A_{\perp}|^2$ & $arg(A_{\parallel})$ & $arg(A_{\perp})$
\\ \hline
CLEO\cite{CLEO-b} &
$0.52\pm0.07\pm0.04$ & $0.16\pm0.08\pm0.04$ & $3.00\pm0.37\pm0.04$ & 
$-0.11\pm 0.46\pm0.03$ \\
CDF\cite{CDF} &
$0.59\pm0.06\pm0.01$ & $0.13^{+0.12}_{-0.09}\pm0.06$ & 
$2.2\pm0.5\pm0.1$ & $-0.6\pm 0.5\pm0.1$ \\
BaBar\cite{BaBar} &
$0.60\pm0.03\pm0.02$ & $0.16\pm0.03\pm0.01$ & $2.50\pm0.20\pm0.08$ &
$-0.17\pm 0.16\pm0.07$ \\ \hline
This measurement &
$0.62\pm0.02\pm0.03$ & $0.19\pm0.02\pm0.03$ &
$2.83\pm0.19\pm0.08$ & $-0.09\pm0.13\pm0.06$ \\ \hline
\end{tabular}
\end{center}
\end{table}

\clearpage
%
%
%

\clearpage
%
%

\clearpage
%
%

\end{document}